\documentclass[prl,twocolumn,showpacs]{revtex4}
\usepackage{epsfig}

\begin{document}
\title{The Critical Exponent of the Fractional Langevin Equation is $\alpha_c\approx 0.402$}
\author{S. Burov, E. Barkai\\
Department of Physics, Bar Ilan University, Ramat-Gan 52900 Israel}

\begin{abstract}
We investigate the dynamical phase diagram of the fractional
Langevin equation and show that critical exponents
mark dynamical
transitions in the behavior of the system.
For a free and harmonically bound particle
the critical exponent $\alpha_c= 0.402\pm 0.002$ marks a transition
to a non-monotonic under-damped phase.
The critical exponent $\alpha_{R}=0.441...$
marks a transition to a resonance phase, when an external oscillating field
drives the system.
Physically, we explain these behaviors using a cage effect, where the
medium induces an elastic type of friction.
Phase diagrams describing the under-damped, the over-damped and critical
frequencies of the fractional oscillator, recently used to model single
protein
experiments, show behaviors vastly different from normal.
\end{abstract}
\maketitle

Anomalous subdiffusion  $\langle x^2\rangle\sim t^\alpha$
with $0<\alpha<1$ is found in diverse physical systems
ranging from charge transport in disordered semiconductors,
and quantum dots,
to relaxation dynamics of proteins
and diffusion of mRNA in live E-coli cells to name a few examples
\cite{Metzler1,COFFEY,Haw,Xie1,Marcus,Kneller,Golding,Sokol}.
On the stochastic level, anomalous diffusion and relaxation is modeled
using the
fractional Fokker-Planck-Kramers equations
\cite{Metzler1,COFFEY,Sokol,Barkai3,Barkai2},
and the fractional Langevin equation
\cite{COFFEY,Xie1,Xie2,Lutz,Pottier,Hangii,Goychuk3,Chaud,Kop}.
For example  recent
single molecule experiments on protein dynamics \cite{Xie1,Xie2}
revealed anomalous Mittag-Leffler  relaxation.
The recorded  anomalous behavior:
the distance between an electron donor
and acceptor pair within the protein, {\em bounded by a harmonic
force},  was successfully
modeled by the fractional Langevin equation (FLE) \cite{Xie1,Xie2}.

 For the standard Langevin equation with white noise,
the well known
under-damped and  over-damped phases and the critical frequency
yield a dynamical phase diagram of the motion.
In this manuscript fractional
over and under-damped behaviors are investigated, which exhibit rich behaviors
vastly different from the normal case.
The most surprising result is
the appearance of a critical exponent for the FLE - $\alpha_c\approx 0.402$.
For $\alpha<\alpha_c$ the over-damped behavior totally disappears,
and the decay to equilibrium is never monotonic.
We will interpret this behavior in terms of a cage effect,
and show that similar critical behaviors are found also
for a particle free of deterministic external forces, and
for the
response of the system to an external time dependent field.
Thus critical exponents control fractional dynamics.
Contrary to expectation,
our  work shows that the transition
between normal diffusion and relaxation $\alpha=1$
to the strongly anomalous case $\alpha\rightarrow 0$,
is not a smooth transition.

The FLE is a generalized Langevin equation \cite{COFFEY,HanREV}
 with a power-law memory kernel
\cite{COFFEY,Xie1,Xie2,Lutz,Pottier,Hangii,Goychuk3,Chaud,Kop}
\begin{equation}
m\frac{d^2 x(t)}{d t^2}=F(x) -\bar{\gamma}\int_0^t\frac{1}{(t-\grave{t})^{\alpha}}
\frac{dx}{d\grave{t}}\,
d\grave{t}+\xi(t) ,
\label{gle1}
\end{equation}
where $\bar{\gamma}>0$
 is a generalized friction constant,
 $F(x)$ is an external force field,
 $0<\alpha<1$ is the fractional exponent and $\xi(t)$ is a stationary,
 fractional Gaussian noise \cite{COFFEY,Mandelb} satisfying
the fluctuation-dissipation relation \cite{COFFEY}
\begin{equation}
\langle\xi(t)\rangle=0,\qquad \langle\xi(t)\xi(\grave{t})\rangle = k_bT\bar{\gamma}|t-\grave{t}|^{-\alpha}.
\label{fluct_dissipate}
\end{equation}
The FLE Eq. (\ref{fluct_dissipate})
can be derived from the Kac-Zwanzig model of a
Brownian particle coupled to an Harmonic bath \cite{Kop}.
And recently the non-trivial fractional
Kramers escape problem for the FLE was solved
\cite{Goychuk3}.
Another convenient way to write Eq. (\ref{gle1}) is
\begin{equation}
m \ddot{x}= F(x) -\overline{\gamma} \Gamma \left(1 - \alpha\right) \frac{d^{\alpha}x}{dt^{\alpha}} + \xi(t),
\label{fle_01}
\end{equation}
where the fractional derivative \cite{Metzler1}  is
defined in Caputo sense
$\frac{d^{\alpha}f(t)}{dt^\alpha}=
\frac{1}{\Gamma(1-\alpha)}
\int_0^t(t-\grave{t})^{-\alpha}(df(\grave{t})/d\grave{t})\,d\grave{t}$.
In recent years fractional calculus was shown to describe many
Physical systems \cite{Metzler1,Sokol}.

{\em The normalized correlation function}
\begin{equation}
C_x(t)=\frac{{\langle}x(t)x(0)\rangle}{{\langle}x(0)^2\rangle},
\label{corrx}
\end{equation}
for a harmonic force field $F(x)= - m \omega^2 x$,
was used to describe single protein dynamics \cite{Xie1,Xie2},
so we will first focus on this experimentally measured quantity.
Thermal initial conditions are used
$\langle\xi(t)x(0)\rangle=0$, $\langle x(0)^2\rangle=k_bT/ m \omega^2$
 and $\langle x(0)v(0)\rangle=0$. We assume that
$\alpha=p/q$, where $q>p>0$ are
integers and $p/q$ is irreducible
(i.e not equal to some other $l/n$ where $l<p$ and $n<q$ are integers).
 Using the convolution theorem
and Eq. (\ref{fle_01}), the Laplace transform of $C_x(t)$ is
\begin{equation}
\hat{C_x}(s)=\frac{s+\gamma s^{\frac{p}{q}-1}}{s^2+\gamma s^{\frac{p}{q}}+ \omega^2}
\label{fractionLap_01}
\end{equation}
where $\gamma=\frac{1}{m}\bar{\gamma}\Gamma(1-\alpha)$.
We  rewrite Eq. (\ref{fractionLap_01}) as
\begin{equation}
\hat{C_x}(s)=\frac{(s+\gamma s^{\frac{p}{q}-1})\hat{Q}(s)}{\hat{P}(s)}
\label{fractionLap_02}
\end{equation}
with
\begin{equation}
\hat{Q}(s)={\left(s^2+\omega^2\right)^q+\left(-1\right)^{q-1}\gamma^qs^{p}\over s^2+\gamma s^{\frac{p}{q}}+\omega^2}
\label{pol_q}
\end{equation}
and
\begin{equation}
\hat{P}(s)=(s^2+\omega^2)^q+(-1)^{q-1}\gamma^qs^{p}.
\label{pol_s}
\end{equation}
By finding the $2q$ zeros of $\hat{P}(a_k)=0$,
 $a_k$ with $k=1,... 2q$,
and using analysis of poles in the complex plane we invert
Eq. (\ref{fractionLap_01}) and obtain an explicit analytical solution for
$C_x(t)$ \cite{Tos1}.
Defining the constants $A_k=1/d\hat{P}(s)/ds|_{s=a_k}$
 and $\tilde{B}_{mj}$ by the expansion
$\hat{Q}(s)(s+\gamma s^{p/q-1})=\sum_{m=0}^{2q-1}\sum_{j=0}^{q-1}\tilde{B}_{mj}s^{m-j/q}$,
the solution is
\begin{equation}
C_{x}(t)=\sum_{m=0}^{2q-1}\sum_{j=0}^{q-1}\sum_{k=1}^{2q}
a_{k}^{m}\tilde{B}_{m\,j}A_{k}t^{\frac{j}{q}}E_{1,1+\frac{j}{q}}(a_kt),
\label{final_sol}
\end{equation}
where $\mathrm{E}_{\eta,\mu}(y)$ is a generalized Mittag-Leffler function
\cite{Metzler1}.
The solution Eq. (\ref{final_sol}) is valid provided that
all the
$a_k$s are distinct, otherwise a critical behavior is found which is
soon discussed.
We see that finding the solution to the problem is equivalent to finding
the zeros of the polynomial $\hat{P}(s)$, once these zeros
are found one can investigate and plot the solution with a program
like Mathematica.
For example for $\alpha=1/2$ Eq. (\ref{final_sol})
reads
\begin{equation}
\begin{array}{l}
C_x(t)=\sum_{k=1}^4
\left(\left(-\gamma^2+\omega^2a_k+a_k^3\right)A_ke^{a_kt} \right. \\
\left.\qquad\qquad
+\gamma\omega^2A_kt^{
\frac{1}{2}}E_{1,\frac{3}{2}}(a_kt)\right).
\end{array}
\label{alpha_halph}
\end{equation}
where $a_k$ are non-identical solutions
of $\hat{P}(s)=(s^2 + \omega^2)^2 - \gamma^2 s=0$.

\begin{figure}
\begin{center}
\includegraphics[width=\columnwidth]{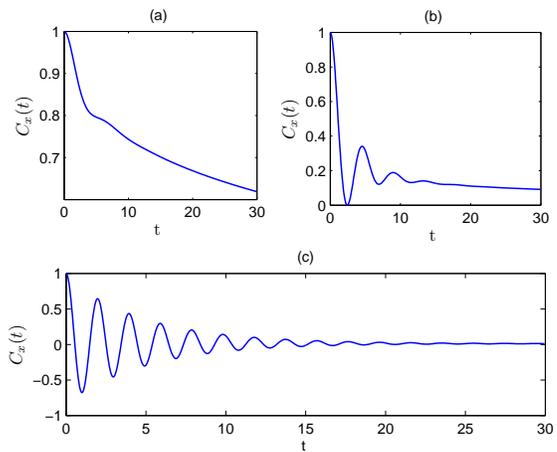}
\end{center}
\caption{ The correlation function
 $C_x(t)$ versus $t$ with $\alpha=1/2$ and natural units  $\gamma=1$.
 Three types of solutions are presented {\bf (a)} $\omega=0.3$ and the correlation function
 decays monotonically.
{\bf (b)}The critical frequency
 $\omega=\omega_z\approx 1.053$ where the transition between motion
with and without zero crossing, namely $C_x(t)=0$  for a single point in
time.
{\bf (c)} For larger frequency  $\omega=3$ we observe non-monotonic decay with
zero crossing.
}
\label{fig1}
    \end{figure}

  The $t\rightarrow\infty$ asymptotic behavior
is found using the exact solution Eq.
(\ref{final_sol})
\begin{equation}
C_x(t)\sim\frac{\gamma}{\omega^2\Gamma(1-\alpha)}t^{-\alpha}
\label{assimptot}
\end{equation}
or with Tauberian theorems \cite{Feller}.
Hence for very long times $C_x(t)$ always decays monotonically.
However for shorter times the picture is very different.
In Fig. \ref{fig1},
the analytical solution for $C_x(t)$ with $\alpha=1/2$
 is plotted for various $\omega$
using natural units $\gamma=1$.
Three types of behaviors exist (i)
Monotonic decay of the solution - Fig. \ref{fig1}(a).
(ii) Non-monotonic decay of the
solution with no zero crossing $C_x(t)\geq 0$ Fig. \ref{fig1}(b).
(iii) When the frequency $\omega$ becomes large
 $C_x(t)$ exhibits non-monotonic decay with zero-crossings Fig. \ref{fig1}(c).
Similar types of  behaviors are found also for other parameter set.

Our goal now is to clarify the phase diagram of the dynamics
of the fractional oscillator. For $\alpha=1$,
the regular damped oscillator, $\omega_c=\gamma/2$ defines the critical
frequency which marks the transition between
over-damped and under-damped motion.
For the fractional oscillator $0<\alpha<1$ the identification
of a single critical frequency is not possible,
and we use
three definitions
for critical frequencies where the dynamics of the system undergoes
a transition in its behavior.
For the normal case $\alpha=1$, the critical frequency
$\omega_c$ is found
when two zeros of $\hat{P}(s)$ coincide,
i.e. an appearance of a pole of a second order for $\hat{C}_x(s)$.
Similarly for the $0<\alpha<1$ case we define $\omega_c$
as a critical point for which not all the zeros of $\hat{P}(s)$
are distinct and as a result for such $\omega_c$
Eq. (\ref{final_sol}) is not valid.
The second approach is to take the minimal frequency
$\omega_z$ at which the solution $C_x(t)$ crosses the zero line.
Finally we will investigate the minimal frequency $\omega_m$
at which  $C_x(t)$ is no longer a monotonically decaying function.
Only for the regular damped oscillator $\alpha=1$, we have $\omega_c=\omega_z=\omega_m$.

\begin{figure}
\begin{center}
\includegraphics[width=\columnwidth]{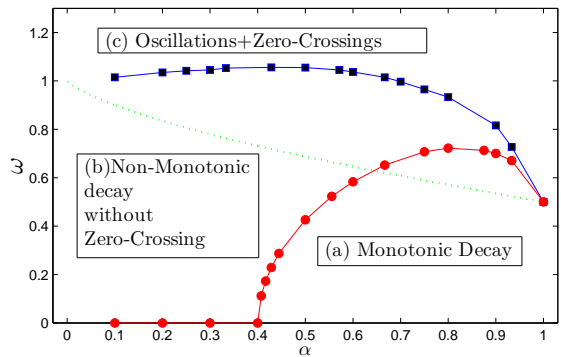}
\end{center}
\caption{
 The phase diagram of the fractional oscillator.
Phase (a) monotonic decay of the correlation function
$C_x(t)$, phase (b) non-monotonic decay without zero-crossing and
(c) oscillations with zero crossings.
 The boundary between (b) and (c) is $\omega_z=\kappa_z(\alpha)$
 (solid line + squares), the boundary between (a) and (b) is
 $\omega_m=\kappa_m(\alpha)$ (solid line + circles).  For
$\alpha<\alpha_c\simeq 0.402$,
the phase of monotonic decay disappears, namely we do not find
over damped behavior.
 The dotted curve is the
 critical line $\omega_c$ given by Eq. (\ref{wc_01}).
All the curves are calculated for $\gamma=1$.
For $\alpha=1$, $\omega_c=\omega_z=\omega_m=\gamma/2$.
}
\label{fig2}
\end{figure}

Eqs. (\ref{fractionLap_01},\ref{final_sol}) are used now
to investigate the phase diagram of the
motion.
By its definition, at the critical frequency $\omega_c$ a pole
of second order in  Eq. (\ref{fractionLap_01})
is found.
It is easy to see that for such $\omega_c$ two conditions
must be satisfied, $\hat{P}(s)=0$ and $d\hat{P}(s)/ds=0$.
These two conditions lead to the following relation for $\omega_c$
\begin{equation}
\omega_c=\frac{1}{2^{\frac{1}{2-\alpha}}}
\sqrt{(2-\alpha)\alpha^{\frac{\alpha}{2-\alpha}}}
\gamma^{\frac{1}{2-\alpha}},
\label{wc_01}
\end{equation}
which is valid only for even $q$ or even $(q+p)$ where $\alpha=p/q$.
For odd $q$ and even $p$, the critical point $\omega_c$ doesn't exist at all.
Similar to Eq. (\ref{wc_01}) we find from dimensional analysis
\begin{equation}
\omega_z=\kappa_z(\alpha)\gamma^{\frac{1}{2-\alpha}}\;\,\qquad
\omega_m=\kappa_m(\alpha)\gamma^{\frac{1}{2-\alpha}}
\label{kappa_01}
\end{equation}
where $\kappa_z(\alpha)$ and $\kappa_m(\alpha)$
depend only on $\alpha$. By investigating the exact analytical solution
Eq. (\ref{final_sol}) for various
$\alpha$ and $\gamma=1$ we obtain the functions
$\kappa_z(\alpha)$ and $\kappa_m(\alpha)$.
The resulting phase diagram of the fractional oscillator
is presented in Fig. \ref{fig2}.

 From the phase diagram Fig. \ref{fig2} a very interesting result
emerges. For $\alpha<\alpha_c \approx 0.4$
the solution is never decaying monotonically.
For such $\alpha$ an oscillatory behavior
is always found even if the frequency of the
binding Harmonic field $\omega\to 0$. In this sense the solution is always
under-damped. Thus we find a critical exponent $\alpha_c$ which marks
a pronounced  transition in the behaviors of the solutions of the
fractional oscillator.
We also note
that for $\alpha$ near
$\alpha_{c}$ we find
$\kappa_m(\alpha)\propto (\alpha-\alpha_{c})^{\frac{1}{2}}$,
which describes the  boundary
between the monotonic and non-monotonic phases.
Note that the phase diagram Fig. \ref{fig2}
also exhibits some expected behaviors:
as we increase $\omega$ we find a critical line above which
the  solutions are non-monotonic and exhibit zero crossing (similar
to usual under-damped behavior)
and for $\alpha=1$ the three
critical frequencies $\omega_c$, $\omega_z$ and $\omega_m$ are identical.

 A physical explanation for the phase diagram is based on
the cage effect.
For small $\alpha$ the  friction force induced by the medium
is not just slowing down the particle but also
causing the particle a rattling motion. To see this
consider the FLE Eq. (\ref{fle_01})
in the limit $\alpha\rightarrow 0$
\begin{equation}
m \ddot{x}+ m \gamma(x-x_0) + m \omega^2 x \approx \xi(t)
\label{regular_oscil}
\end{equation}
where $x_0$ is the initial condition.
Eq. (\ref{regular_oscil}) describes harmonic motion
{\em and the ``friction" $\gamma$ in this
$\alpha \to 0$ limit yields an elastic
harmonic
force}. In this sense the medium is binding the particle preventing
diffusion but forcing oscillations.
In the opposite limit of $\alpha\to 1$
\begin{equation}
m \ddot{x}+ m \gamma\dot{x} +m  \omega^2 x \approx \xi(t) \qquad\mbox{when}\;\alpha\rightarrow 1,
\label{damped_oscil}
\end{equation}
the usual  damped oscillator with noise is found.
 So from Eq. (\ref{regular_oscil})
for $\alpha\rightarrow 0$ an oscillating behavior is expected,
{\em even when $\omega \to 0$}
which can be explained by the rattling motion of a particle
in the cage formed by the surrounding particles.
This behavior manifests itself in the non-monotonic oscillating solution
we have found in our phase diagram Fig. \ref{fig2}
when $\alpha \to 0$.
Our finding that $\alpha_c$ marks a
non smooth transition between normal friction $\alpha\to 1$ and
elastic friction $\alpha \to 0$ is certainly
a surprising result
which could not be anticipated without our mathematical analysis.

To find accurate values of $\alpha_c$
 we used also a method based on Bernstein theorem \cite{Feller}.
According to the theorem if and only if a function
$f(t)$ is positive then for any integer $n$,
 $0\leq(-1)^n(d^n\hat{f}(s)/ds^n)$, where $\hat{f}(s)$
is the Laplace pair of $f(t)$.
To see when $C_x(t)$ is decaying monotonically,
we investigate the positivity of its derivative.
Using Bernstein's theorem
inspecting the $n=150$ order derivatives
gives the exact lower bound  $\alpha_{c}\geq 0.394$,
while the method using exact solution for $C_x(t)$ gives
$\alpha_c=0.402 \pm 0.002$.
Further we can show that the critical exponent is
not equal to $0.4$, and hence it seems a non trivial number.
We now show that the critical exponent $\alpha_c$
is important for other physical quantities.

{\em The mean square displacement $\langle x^2(t)\rangle$}
for the force free particle
$F(x)=0$ is now investigated. We consider
averages over the noise and thermal initial
conditions $\langle v^2\rangle|_{t=0}=k_bT/m$. The analytical
solution for this case was found already in \cite{Barkai2,Lutz}
and is given by
$\langle x^2(t)\rangle=2(k_bT/m)t^2E_{2-\alpha,3}(-\gamma t^{2-\alpha})$.
We find that for $\alpha<\alpha_c$
$\langle x^2(t)\rangle$ is non-monotonic,
while for $\alpha>\alpha_c$ it is monotonic.
The non-monotonic behavior of
$\langle x^2(t)\rangle$ for small $\alpha$ is
a manifestation of a cage effect. Our finding
that the same critical exponent $\alpha_c$
describes both the phase diagram of the fractional
Harmonic oscillator and also
the fluctuations of a particle free of an external force field,
emphasizes the generality of our results.
Although not explored here in detail,
the same $\alpha_c$ is likely to describe motion in
any binding potential field, since it is the medium
inducing the oscillations.
We note that  our results can be derived
also from  the fractional Kramers equation
\cite{Barkai2,remark2} and hence are not
limited to the FLE.

\begin{figure}
\begin{center}
\includegraphics[width=\columnwidth]{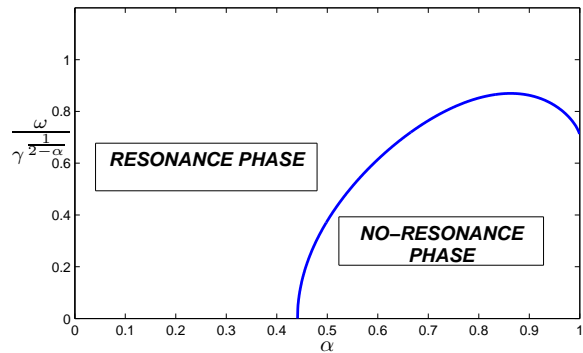}
\end{center}
\caption{
Phase diagram of the response of the system
to a oscillating time dependent force field.
Two simple behaviors are found either a resonance exists
or not.
For $\alpha<\alpha_{R}=0.441...$ a resonance exists for any
binding harmonic field and any friction.
}
\label{fig3}
\end{figure}

{\em The response of a system} to an oscillating time dependent field
naturally leads to the phenomena of resonances, when the frequency
of the external field matches a natural frequency of the system.
The response of sub-diffusing systems to such time
dependent fields was the subject of intensive research
\cite{Barbi,Sokolov2,Goychuk1}. In particular fractional approach
to sub-diffusion naturally leads to anomalous response functions
commonly found in many systems e.g. the Cole-Cole relaxation
\cite{Barkai3,Goychuk2}.
Here we investigate the response of a particle in a harmonic force
field, as found in the experiments \cite{Xie1,Xie2}, to a time dependent force
$F_0\cos(\Omega t)$ using Eq. (\ref{fle_01}).
In the long time limit, using the stationarity of the process we find
\begin{equation}
\langle x \rangle
\sim \frac{\mathcal{F}_0}{m} R(\Omega) \cos\left[\Omega t +\theta(\Omega)\right]
\qquad t\rightarrow \infty,
\label{x_oscill}
\end{equation}
where the response function is
\begin{equation}
R(\Omega)= \left| \left[ \omega^2 + \gamma\left( - i \Omega\right)^\alpha - \Omega^2\right ]^{-1}  \right|,
\label{EqRes}
\end{equation}
which was derived already in \cite{Coffey1} using a different model.
As well known for the normal case $\alpha=1$ a resonance is found when the
friction is weak, namely  a peak
in $R(\Omega)$ is observed  when $\omega > \gamma/\sqrt{2}$.
On the other hand
if the
damping is strong $\omega<\gamma/\sqrt{2}$ the peak is not found.
The simple phase diagram describing the response of the fractional system
is shown in Fig. \ref{fig3}, where a resonance phase means a phase with a
peak in the response function.
For $\alpha$ smaller than a critical exponent $\alpha_{R}=0.441...$
one always detects a resonance, even if $\omega \to 0$.
This is clearly a manifestation of the cage effect and it clearly
shows that the critical exponent $\alpha_{R}$ marks a
dynamical transition of the response
of the system to an external
force field.

 Many works consider the  over-damped approximation
to anomalous diffusion, which means that Newton's acceleration
term is neglected, $m=0$ in Eq.
(\ref{gle1}). This approximation must be used with care.
As we showed when $\alpha<\alpha_{c}$ the exact solution of the FLE
always exhibits a non-monotonic
decay of $C_x(t)$,
while the over-damped approximation gives a monotonic decay.
And according to our results when $\alpha<\alpha_{R}$ resonances always
appear in the response of the system to an external oscillating
field, which are not found with the over-damped approximation.
In this sense the
over-damped approximation fails when $\alpha$ is smaller
than the critical exponents of the system.
This is in complete contrast to the usual Langevin equation with
white noise, where the over-damped approximation captures the main
features of the dynamics.
This striking difference between the FLE and the usual Langevin equation
stems from the fact that for the FLE the dissipative
memory induces oscillations, as mentioned.

 In conclusion,
critical exponents mark sharp transitions in the behaviors of systems
with fractional dynamics.
The critical exponents describe a wide range of physical behaviors:
the
correlation function $C_x(t)$ of a particle bounded by a Harmonic field,
the mean square
displacement of the free particle, and the response of the
system to an external oscillating field. Thus these critical exponents
are clearly very important and
general in the description of the anomalous kinetics.
The phase diagrams we obtained are
related to a cage effect,
where for small enough $\alpha$ the medium induces oscillations
in the dynamics of the particle.
Thus the fractional dynamics is
profoundly different from the dynamics described
by the Langevin equation with white
noise.

{\bf Acknowledgment} This work was supported by the Israel Science
Foundation.


\begin{thebibliography}{999}
\bibitem{Tos1} In our analysis we make use of the fact that
$\hat{Q}(s)$ is a polynomial in $s^{1/q}$ (and hence we can perform
the inverse Laplace transform)
since
any solution of $y^{2q}+Ay^{p}+w^2=0$ is also a
solution of $(y^{2q}+w^2)^q+(-1)^{q-1}A^q{y^{qp}}=0$.
\bibitem{remark2} The fractional Kramers-Fokker-Planck framework
and the
FLE
are non-identical and competing
stochastic approaches, however
for $C_x(t)$ and $\langle x^2 \rangle$
they give the same results.
\end{thebibliography}
\end{document}